\documentclass[journal=jacsat,manuscript=article]{achemso}
\usepackage{graphicx}
\usepackage{subcaption}
\usepackage[version=3]{mhchem}
\usepackage{amsmath}
\usepackage{amsthm}
\usepackage{nicefrac}
\usepackage{amssymb}
\usepackage{amsfonts}
\usepackage{mathtools}
\usepackage{siunitx}
\usepackage{blindtext}
\usepackage{comment}
\usepackage{xcolor}
\usepackage{xr}
\usepackage{lineno}
\usepackage[colorlinks]{hyperref}
\usepackage{float}

\author{Carlos David Gonz\'alez-G\'omez}
\affiliation{Nanoparticles Trapping Laboratory, Department of Applied Physics, Universidad de Granada, 18071 Granada, Spain}
\alsoaffiliation{Universidad de M\'alaga, Departamento de Física Aplicada II, 29071 M\'alaga, Spain}

\author{Ra\'ul A. Rica}\email{rul@ugr.es}
\affiliation{Nanoparticles Trapping Laboratory, Department of Applied Physics, Universidad de Granada, 18071 Granada, Spain}
\alsoaffiliation{Universidad de Granada, Research Unit “Modeling Nature” (MNat), 18071 Granada, Spain}

\author{Emilio Ruiz-Reina}\email{eruizr@uma.es}
\affiliation{Institute Carlos I for Theoretical and Computational Physics (iC1), Departamento de Física Aplicada II, Universidad de M\'alaga, 29071 M\'alaga, Spain}

\title{Electro-Thermo-Plasmonic Flow in Gold Nanoparticle Suspensions: Nonlinear Flow Velocity Dependence with Aggregates Concentration}

\begin{document}

\begin{abstract}

Efficient mixing and pumping of liquids at the microscale is a technology that is still to be optimized. The combination of an AC electric field with a small temperature gradient leads to strong electro-thermal flow that can be used for multiple purposes. Combining simulations and experiments, an analysis of the performance of electro-thermal flow is provided when the temperature gradient is generated by illuminating plasmonic nanoparticles in suspension with a near-resonance focused laser. Fluid flow is measured tracking the velocity of fluorescent tracer microparticles in suspension as a function of electric field, laser power and concentration of plasmonic particles. Among other results, a non-linear relation is found between the velocity of the fluid and particle concentration, what is justified in terms of multiple scattering-absorption events, involving aggregates and individual particles, that lead to enhanced absorption when the concentration is raised. Simulations provide a description of the phenomenon that is compatible with experiments, and constitute a way to understand and estimate the absorption and scattering cross-sections of dispersed particles and/or aggregates. Comparison of experiments and simulations suggest that the gold nanoparticles are aggregated forming clusters of about 5-9 particles, but no information about their structure cannot be obtained without further theoretical and experimental developments. This nonlinear behavior could be useful to get very high ETP velocities by inducing some controlled aggregation of the particles.

\end{abstract}

\maketitle

\section{Introduction}

There are many applications based on lab-on-a-chip technologies that involve transport and mixing of fluids inside microchannels \cite{Hatch2004Diffusion,Gervais2006Transport,Squires2008Making}, but fulfilling their demands requires further developments. One hurdle to be overcome comes from the fact that mixing essentially occurs by diffusion, which is inherently slow \cite{squires2005microfluidics}. Moreover, laminar flow is typically present at the microscale due to the small Reynolds numbers attained, so convection has to be actively forced in order to speed up mixing. This is particularly important in the case of applications based on surface sensors \cite{Sheehan2005Detection,Protiva2011Enhancing} or where mixing between different species is intended\cite{marschewski2015mixing}. Active pumping is also required in many applications, and the standard approach involves the use of syringe pumps that create pressure-driven flows \cite{Lake2017Low,Richardson2009Experimental,Laser2004Review,Iverson2008Recent}. However, miniaturization and portability clearly require the use of alternative approaches \cite{SCHONBERGER2016235,arshavsky2020lab}. 

Convection can be triggered by strong temperature gradients. These can be obtained e.g. via light-to-heat conversion with plasmonic nanostructures deposited on a substrate\cite{Donner2011Plasmon,Roxworthy2014Understanding,Ciraulo2021Long,Baffou2017Thermoplasmonics,kotnala2019overcoming, Chen2021Heat,acimovic2014LSPR,baffou2013thermo,Roxworthy2014Understanding}. Likewise, electrokinetic techniques have been extensively developed both to induce convection and pump flow inside microchannels \cite{ramos1999ac2,zeng2001fabrication,olesen2006ac,Chang2007Electrokinetik,garcia2008traveling,bazant2010induced}. Nevertheless, these approaches present some drawbacks that prevent widespread application. In the case of plasmonic structures, the generation of strong flows requires somehow high temperature increments (tens of K), what can compromise their use with analytes of biomedical interest,  since the most of them are very sensible to temperature variations. On the other hand, electrokinetic pumps require high electric fields for their efficient operation \cite{wang2009electroosmotic,manshadi2020induced,weiss2022chip}. Moreover, both approaches require costly microfabrication techniques.   
 
Alternatively, electro-thermal flow can be efficiently generated combining a small temperature gradient with an AC electric field \cite{Ramos1998AC,green2001electrothermally,castellanos2003electrohydrodynamics}. A plasmonic nanostructure used as a heat source in combination with an AC electric field has been shown to provide versatility in the generation of convective flow inside microchannels thanks to the electro-thermo-plasmonic (ETP) effect \cite{ndukaife2014photothermal,Ndukaife2016Long,garcia2018overcoming,hong2021electrothermoplasmonic}. This approach is attractive since the ETP effect has been shown to provide significant flow with low temperature increments and moderate electric fields. Interestingly, the flow thus obtained can be even engineered to create stagnation points that are able to trap molecules and small nanoparticles in a region where no heating is produced \cite{Hong2020Stand}. 

In this work, we explore a novel configuration for ETP flow where heating is achieved thanks to the laser absorption of plasmonic gold nanoparticles dispersed in the fluid, which constitutes an interesting alternative to the deposition of gold nanostructures on a substrate\cite{garcia2018overcoming,Ciraulo2021Long}. This platform is used to better understand and characterize the dynamics of ETP flow. In our experiments, we present measurements of the flow field obtained as a function of the electric field, the laser power, and the concentration of dispersed gold nanoparticles (AuNPs). We compare our results with simulations performed with COMSOL Multiphysics, providing an in-depth understanding of the mechanism behind ETP flow.

The paper is organized as follows. We firstly introduce ETP fundamentals, discussing the particular structure of ETP flow obtained in our configuration aided by computer simulations. Further, we present experimental results of the flow field obtained under different experimental situations, comparing these with simulations.

\section{Overview of electro-thermo-plasmonic flow}

The electro-thermal effect is a flow generation technique that combines a gradient of temperature $\nabla T(\textbf{r})$ with an applied AC electric field $\textbf{E}$. In this situation, the electric field exerts a body force on the fluid due to gradients in electric permittivity and conductivity concomitant with the temperature gradient. The driving electrothermal force is \cite{Ramos1998AC}: 
\begin{equation}
\label{eq: ETP}
\textbf{F}_{\text{ET}}(\textbf{r})=\frac{1}2\text{Re}\left[\frac{\varepsilon(\alpha-\beta)}{1+i\omega\tau}(\nabla T(\textbf{r})\cdot\textbf{E})\textbf{E}^*-\frac{1}2\varepsilon \alpha|\textbf{E}|^2 \nabla T(\textbf{r})\right]
\end{equation}
where $\alpha=(1^{}/{\varepsilon})(\partial \varepsilon^{}/{\partial T})$, $\beta=(1^{}/{\sigma})(\partial \sigma^{}/{\partial T})$, and $\sigma$ and $\varepsilon$ are the conductivity and permittivity of the solution, respectively, at the angular frequency $\omega$ of the electric field. Also, $\tau=\varepsilon^{}/{\sigma}$ is the charge relaxation time of the solution. The second term in this force expression starts to dominate over the first one above the MHz range, determined by the inverse of the relaxation time of the solution \cite{Ramos1998AC, garcia2018overcoming}. 

\begin{figure}[ht!]
	\centering
	\includegraphics[width=15cm]{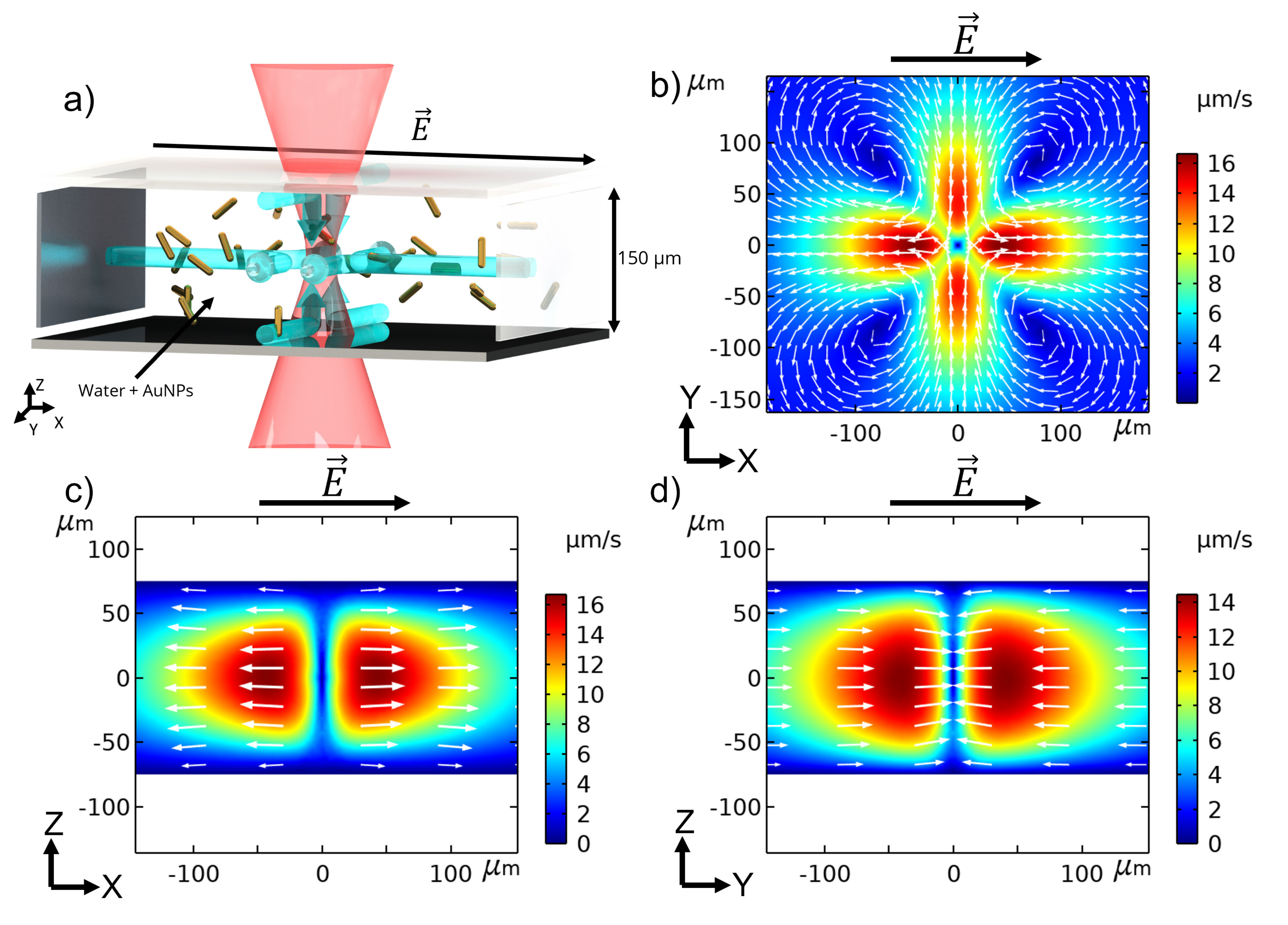}
	\caption{\textbf{Schematic representation of the problem and simulations of the fluid-velocity field}. (a) A microfluidic chamber is fabricated by sandwiching two aluminum-foil electrodes with two glass coverslips. An infrared laser ($\lambda = 980 \text{nm}$) is focused by an objective microscope (20x, NA = 0.4), and excites off-resonance the surface plasmon of gold nanoparticles (AuNPs) uniformly dispersed in the bulk. An AC electric field generated in the bulk by feeding the electrodes with an AC voltage signal ($f=100$kHz) between two electrodes that apply an oscillating electric field with intensity in the range 10-50 kV/m. Panels (b), (c) and (d) show the simulated flow fields in the XY, XZ and YZ planes, respectively. In the simulations, the AuNP concentration is $2.6\cdot10^{12} \text{particles}/{\text{cm}^3}$, the incident laser power is 30 mW and the applied electric field is 30 V/mm. }
	\label{fig:ETPSimul}
\end{figure}

The experimental configuration we have used is depicted in Fig.\ref{fig:ETPSimul}a). A microfluidic chamber is constructed by sandwiching two electrodes made of aluminum foil (150 $\pm$ 20 $\mu$m thickness) with two glass coverslips. The typical distance between electrodes was 1 mm. In our experiment, the temperature gradient is obtained via resonant excitation of surface plasmons in gold nanoparticles that are dispersed in the fluid \cite{baffou2013thermo}. This is different from previous approaches, where ETP flow was generated by excitation of nanostructures deposited on a surface \cite{Ndukaife2016Long,garcia2018overcoming}. Therefore, the geometry of the convection cells we obtain is different, since in our case the electric field is perpendicular to the laser beam (see Fig.\ref{fig:ETPSimul}a)), while in previous works the laser and the electric field were always parallel \cite{Ndukaife2016Long,garcia2018overcoming,hong2021electrothermoplasmonic}. Experiments and numerical simulations based on finite element analysis and performed with COMSOL Multiphysics (see Methods section) reveal that the flow field features a quadrupolar structure in the focal plane (XY). The liquid flows inwards (outwards) from (to) the focus in the directions perpendicular (parallel) to the electric field, respectively (see Fig. \ref{fig:ETPSimul}). Interestingly, the flow is significant, leading to vortexes moving at tens of $\mu\rm m/s$ over distances hundreds of microns apart from the focus, what can be useful in mixing applications.  

In order to experimentally study the ETP flow field in the perpendicular configuration, we prepared suspensions of gold nanoparticles and fluorescent microparticles. The former are used to generate the temperature gradient, while the later serve as flow tracers. The flow tracers will move with a velocity $\bf{v}=\bf{u}+\bf{F}/\gamma$, where $\bf{u}$ is the velocity of the flow, $\bf{F}$ is the sum of the external forces acting on the particles and $\gamma=6\pi\eta r$ is the Stokes friction coefficient, which depends on the fluid viscosity $\eta$ and the tracer particle radius $r$. Since the electric field is AC and with negligible gradients in our electrode configuration, we can safely ignore additional forces on the particles and identify the velocity of the fluid with that of the tracers. 

\begin{figure}[ht!]
	\centering
	\includegraphics[width=16.5cm]{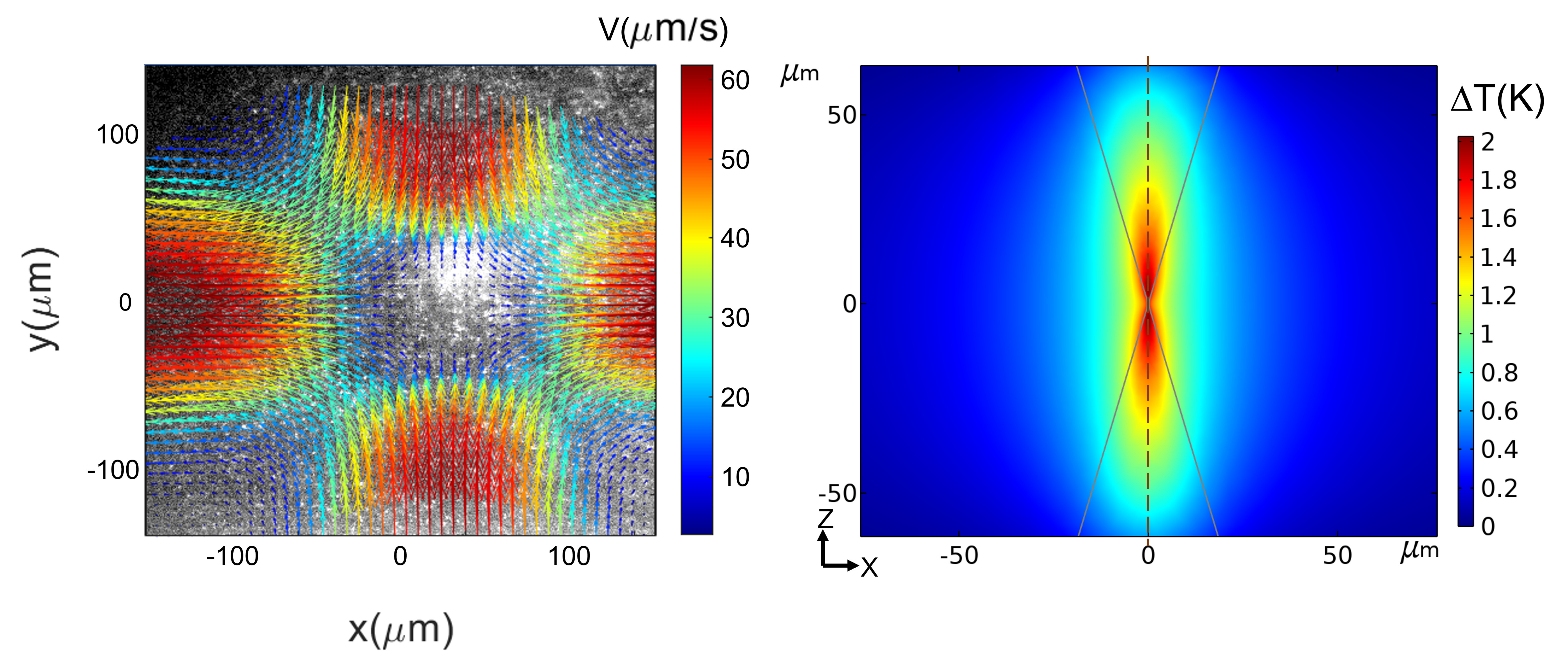}
	\caption{Left: Experimental flow field. The AuNP concentration is $4.96\cdot10^{12} \text{particles}^{}/{\text{cm}^3}$, the incident laser power is 30 mW and the applied electric field is 30 kV/m. Right: Simulated temperature increment corresponding to the experimental condition shown in left. Grey line corresponds to laser beam profile.}
	\label{fig:EXPFF}
\end{figure}

Videos of the flow motion were recorded on a CMOS camera and analyzed with PIVLab\cite{thielicke2014pivlab,thielicke2021particle} to produce experimental flow fields like the one shown in the left panel of Fig.\ref{fig:EXPFF}. As it can be seen, we observe the quadrupolar structure predicted by simulations (compare with Fig.\ref{fig:ETPSimul}). In this case, flow velocities as high as 60 $\mu$m/s are obtained with moderately low electric fields (30 V/mm). Even if our experimental setup does not allow us to measure the temperature of the fluid, comparison with simulations provides an estimate. The right panel in Fig.\ref{fig:EXPFF} also depicts the temperature field that matches the experiment in the left panel of Fig.\ref{fig:EXPFF}. The simulation considers heat transfer and the intensity distribution of the focused Gaussian beam\cite{Rodriguez2020Heat} (see Methods section for a detailed description). As it can be seen, the temperature field presents two lobes before and after the focus where the fluid temperature is raised due to heat generation upon plasmonic excitation. Interestingly, simulations predict that just a 2 K temperature increment at the focus is enough to develop convection cells with the observed velocities.

\section{Results and discussion}

\subsection{Experimental results on electrothermoplasmonic flow}

In this section, we analyse experimentally the influence of the different parameters in the ETP flow, by evaluating the maximum velocity observed in each experiment. Similar results are obtained if the average velocity is considered instead of the maximum velocity.

We performed multiple experiments in order to evaluate the dependencies of the flow field with the electric field, laser power and gold nanoparticles concentration. The frequency of the AC field in these studies was kept at 100~kHz. This choice is based on the fact that the electrothermal flow has been shown to be independent from it in the range [10~kHz-1~MHz]\cite{castellanos2003electrohydrodynamics,garcia2018overcoming}. This is the case when the condition $\varepsilon\omega/\sigma<<1$ is fulfilled, as it is the case for suspensions made with CO$_2$ saturated water with no electrolytes added. Likewise, ETP flow is expected to be the main driving force in such regime, given the size of our chamber and the used electric field. Notice that other effects like AC electroosmosis can dominate fluid flow if the distance between the walls is decreased\cite{castellanos2003electrohydrodynamics}. In simulations, AC electroosmosis was not included, but buoyancy was taken into account (see Methods). The lower frequency bound safely excludes electrolysis and polarization effects on the electrodes.

\begin{figure}[ht!]
	\centering
	\includegraphics[width=16 cm]{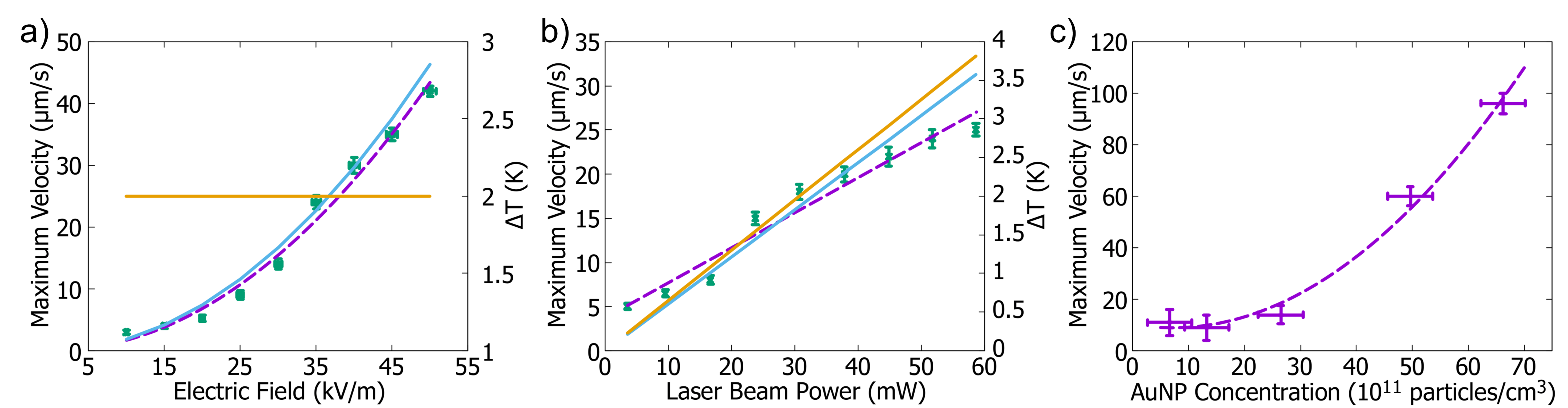}
	\caption{a) Experimental data (\textit{dots}), experimental fit (\textit{dashed line}) and simulation results (\textit{blue solid line}) of maximum flow velocity for an incident laser power of 30 mW and different applied electric fields. The orange solid line is maximum temperature increment, as obtained from simulations. b) Experimental data (\textit{dots}), experimental fit (\textit{dashed line}) and simulation results (\textit{blue solid line}) of maximum flow velocity for an applied electric field of 30 kV/m and different incident laser power. The orange solid line is maximum temperature increment, as obtained from simulations.c) Maximum flow velocity for different AuNP concentrations and an applied electric field of 30 kV/m and an incident laser power of 30 mW.}
	\label{fig:EF_LP_Con}
\end{figure}

In order to characterize the ETP flow, we have performed experiments analysing the influence of electric field magnitude, excitation laser power and particle concentration. The results are shown in Figs.\ref{fig:EF_LP_Con} a), b) and c). The particle concentration value for the data in Figs.\ref{fig:EXPFF}, \ref{fig:EF_LP_Con}a), \ref{fig:EF_LP_Con}b) was the same as in Fig.\ref{fig:ETPSimul}. We first analyze the effect of the applied electric field. Fig.\ref{fig:EF_LP_Con} a) shows that the maximum velocity scales quadratically on the applied electric field, as expected for ET flow (see Eq.\ref{eq: ETP}). Together with the experiments, we present the results of the simulations. For this aim, the only unknown input is the absorption cross-section of the particles in suspension, which we used as a fitting parameter. Since our gold nanoparticles have tendency to aggregate \cite{Arenas2018Electro}, a direct theoretical estimation of this value from simulations is not feasible, as we discuss later. In order to get a consistent analysis, we performed all the simulations with the same value of the absorption cross-section, which is considered a fitting parameter and will be discussed later. Therefore, all ETP flow simulations have been performed using a fitted value of 2.7·$10^{-17} \text{m}^2$ for the absorption cross-section. With this value, the agreement between the predictions of the simulations and the experimental results is adequate in all cases. 

Likewise, in Fig.\ref{fig:EF_LP_Con} b), where the applied electric field is fixed, the relationship between the flow field and the generated heat is linear as expected from Eq.\ref{eq: ETP}. The temperature increment raises with the infrared laser power, leading to an ever increasing velocity. Deviations from linearity are likely due to variability in the construction of the flow chambers.

Finally, we characterized the influence of the concentration of AuNPs on the flow velocity. We performed experiments with five different concentrations at fixed electric field and laser intensity. The results are shown in Fig.\ref{fig:EF_LP_Con} c), where we observe that the maximum velocity has a quadratic dependence with particle concentration. Velocities as high as 100 $\mu$m/s are observed at moderate values of electric field and laser power, and we estimate that up to 0.5 mm/s could be achieved with the maximum values of electric field and laser power we tested in our set of experiments. However, we could not measure adequately this value due to the limited acquisition speed of our camera (25.8 fps for the full sensor). 

The origin of this nonlinearity is unclear, but we hypothesize that it has to do with the way photons are scattered and absorbed by the AuNPs. A proper modelling of the propagation of the laser beam through the sample would involve the solution of an equation of energy transfer, which is out of the scope of this work \cite{Hogan2014NP,mayerhofer2020bouguer}. However, some considerations can be made in order to get information from this result.

For purely absorbing media, the Beer-Lambert law predicts that the absorption of light should increase linearly with concentration. In this situation, increasing the concentration of AuNPs would have a similar effect on the velocity as the increase of laser power, and the dependence should be linear, which is not the case.

In order to account for the observed non-linearity, we need to consider multiple scattering events \cite{Hogan2014NP}. A photon that interacts with a NP can be either scattered or absorbed, with a probability that depends on the respective cross-section. When the concentration increases, a scattered photon is likely to interact with neighbouring particles, increasing its probability to be absorbed. Scattering-absorption events are likely to enhance the absorbance of the suspension as particle concentration increases beyond the linear regime when the scattering and absorption cross-sections are comparable in magnitude\cite{Hogan2014NP}. As we will show later, at the particle concentration considered, this could happen when there is certain aggregation present in the suspension. In order to explore this possibility, we need to study the scattering and absorption cross-sections of the AuNPs used in our experiments, both as individual particles or being part of aggregates. Since we do not have a direct method to measure them, we performed numerical simulations, as described in the next section.

\subsection{Absorption and Scattering Cross-Sections}

The ETP flow is generated by the combination of an applied AC electric field and a thermal gradient, being the latter one a direct consequence of the incident laser energy absorption and scattering by the AuNPs (both as individual particles and as part of aggregates). This means that, in order to analyse the ETP flow behavior, we need as a first step the study of the absorption and scattering cross-sections. The heat power generated inside the AuNPs appears as an input to the calculation of the temperature and flow fields by solving the coupled heat transfer and momentum balance equations.

\begin{figure}[htb]
	\centering
	\includegraphics[width=15cm]{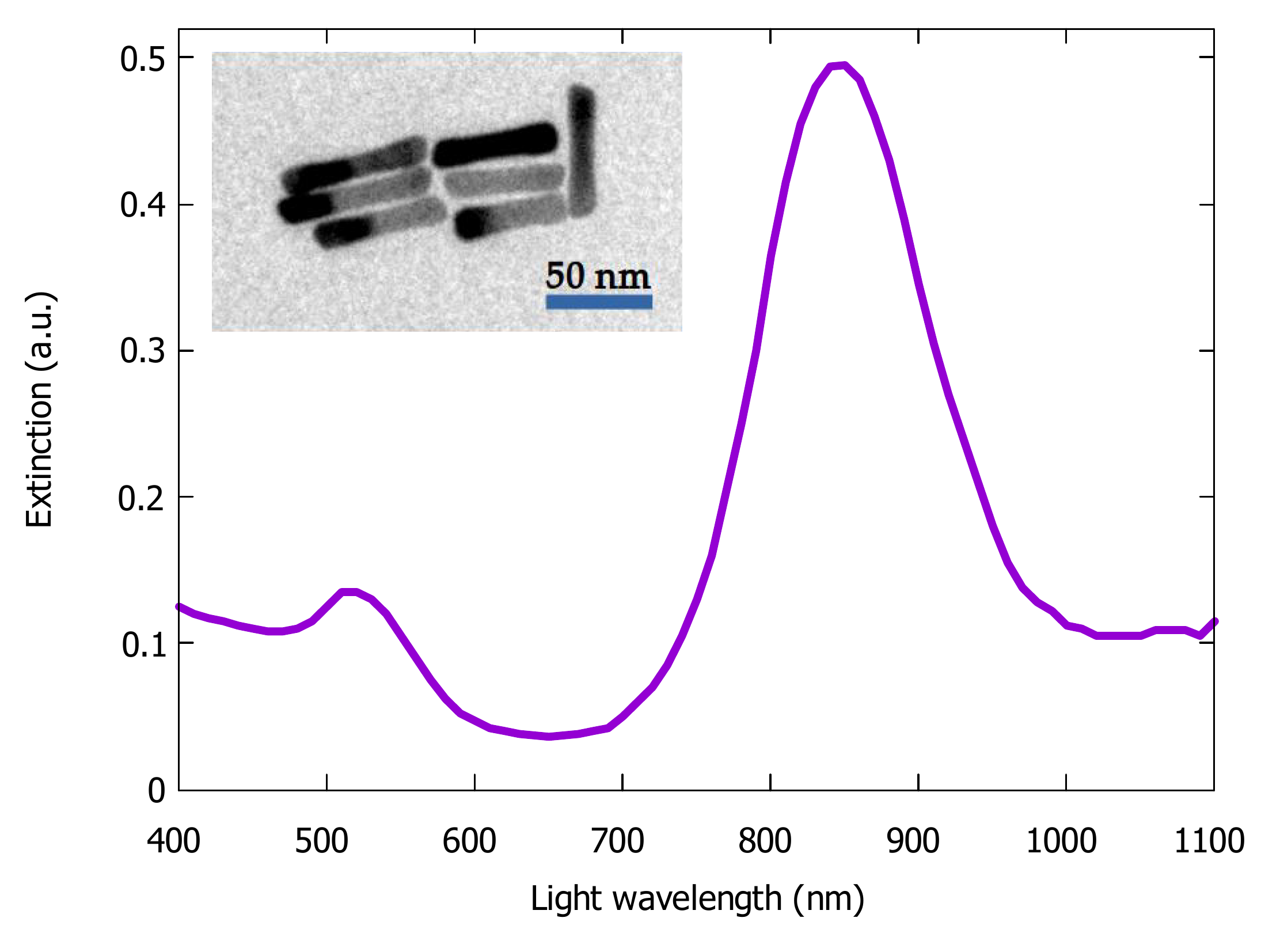}
	\caption{Extinction (scattering + absorption) spectrum of the gold nanorods suspension \cite{Arenas2018Electro}. The inset is a high-resolution TEM picture of gold nanoparticles (AuNPs) sample.}
	\label{fig:AbsAu}
\end{figure}

As can be seen in the experimental extinction curve, Fig.\ref{fig:AbsAu}, suspensions of these particles have a plasmon resonance around 850 nm. The laser wavelength used in the ETP experiments was 980 nm, away from the resonance in order to avoid excessive heating and thus generate a moderate temperature gradient.

\begin{figure}[ht!]
	\centering
	\includegraphics[width=15cm]{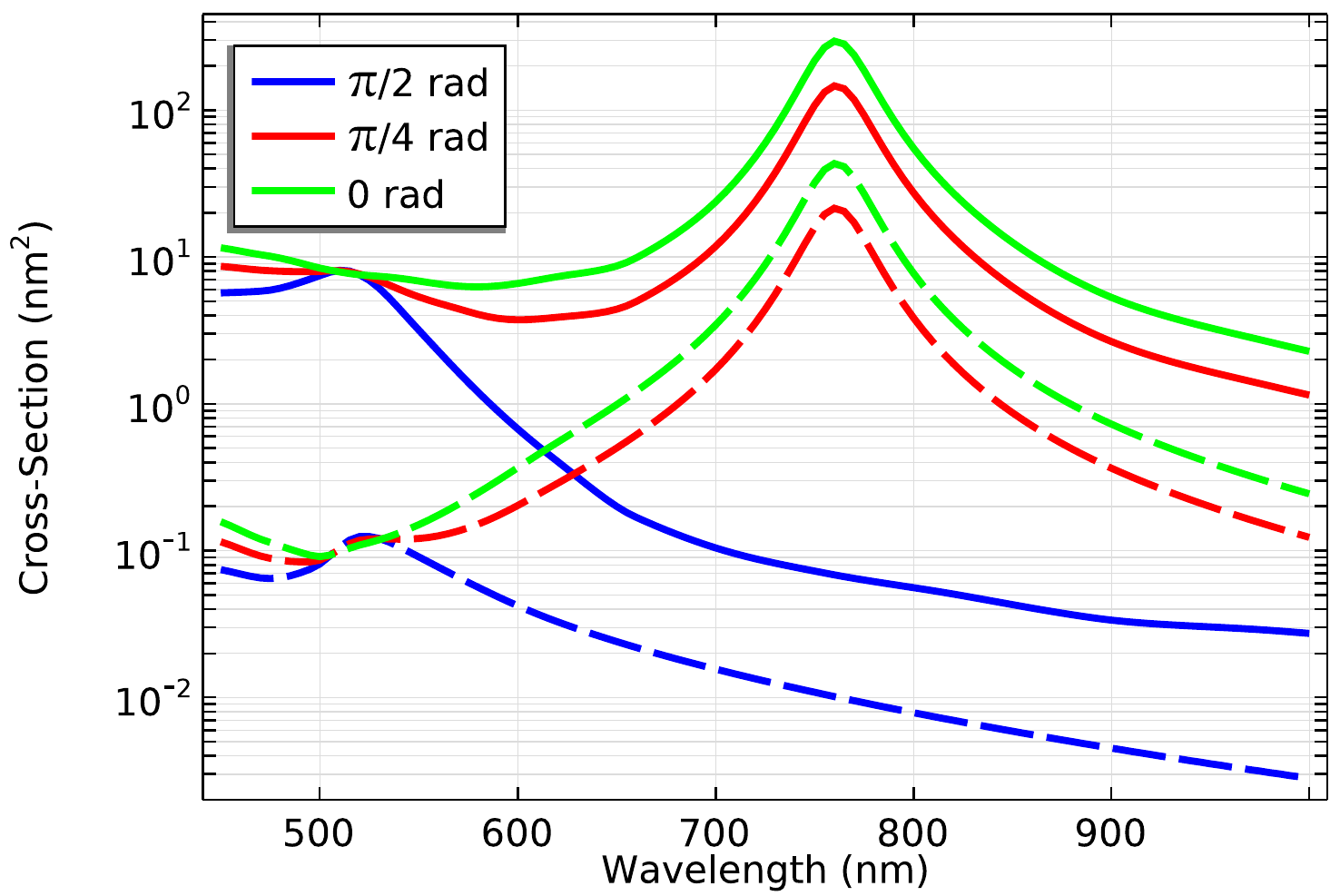}
	\caption{Absorption (solid) and scattering (dashed) simulated cross-section spectra of a single gold nanoparticle for different angles $\theta$ between the incident electric field polarization and the AuNP major axis.}
	\label{fig:CrossSectionWaveLengthSim}
\end{figure}

\begin{figure}[ht!]
	\centering
    \includegraphics[width=15cm]{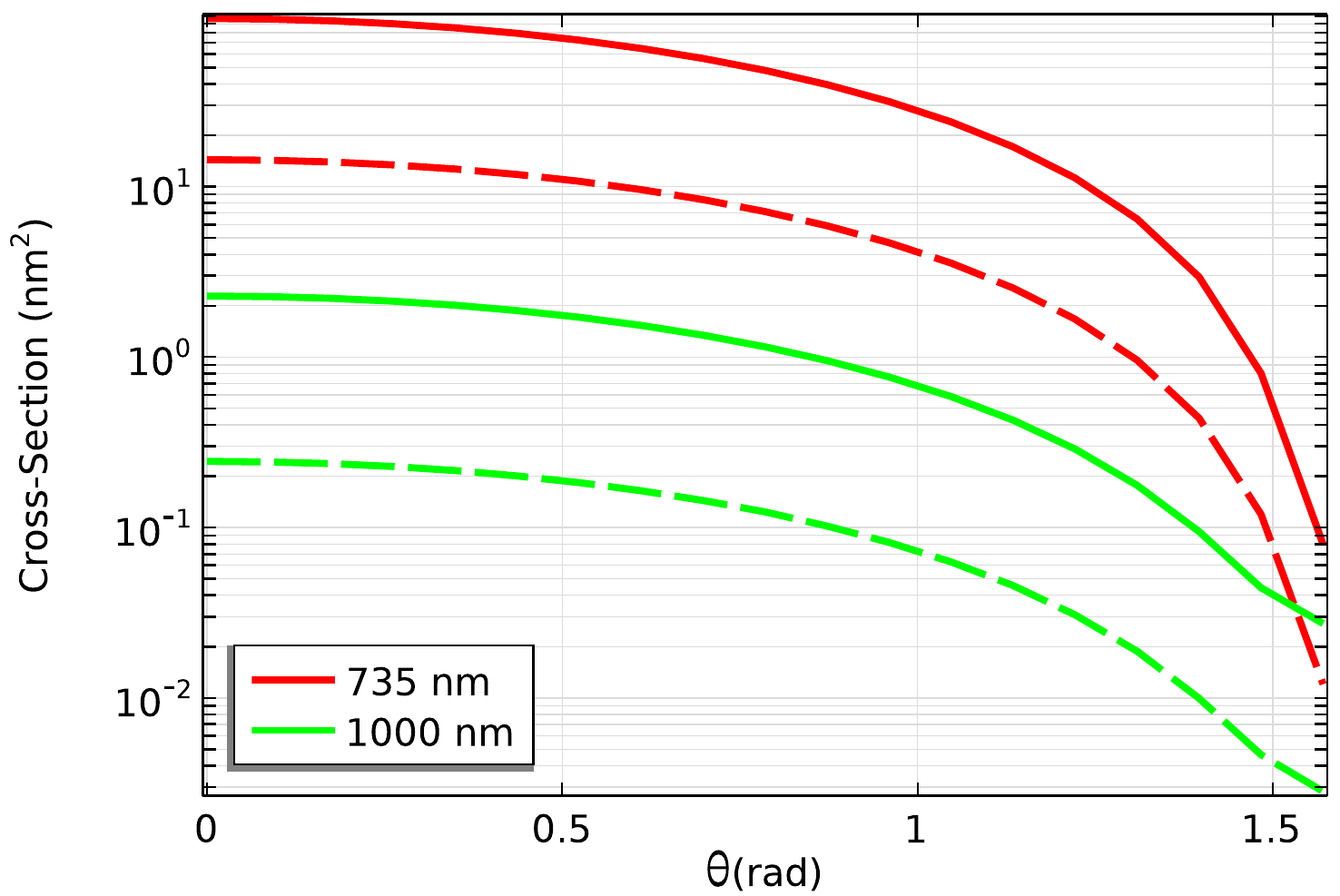}
	\caption{Absorption (solid) and scattering (dashed) simulated cross-sections of a single gold nanoparticle as a function of the angle $\theta$ between the incident electric field polarization and the AuNP major axis.}
	\label{fig:CrossSectionWaveTheta}
\end{figure}

We have performed electromagnetic wave numerical simulations in the wavelength domain, using the finite-element analysis COMSOL Multiphysics platform to calculate the absorption and scattering cross-sections of an individual AuNP, which is modeled as a cylinder with rounded bases, 56 nm length and 16 nm diameter\cite{Arenas2018Electro}. The numerical results obtained are shown in Figs.\ref{fig:CrossSectionWaveLengthSim}-\ref{fig:CrossSectionWaveTheta}.

The highest calculated plasmon resonance peak is located at a frequency about 780 nm. This result is not in agreement with the experimental absorption spectra of our nanorods suspension, see Fig.\ref{fig:AbsAu}, while it fits previous results obtained with similar particles \cite{He2010Scattering}. Experimentally, the main resonance peak appears around 850 nm wavelength and is much broader. This disagreement indicates that our suspension cannot be constituted by individual nanoparticles well separated within the aqueous medium. The reason for this discrepancy lies in the presence of a certain level of particle aggregation, as will be discussed later.

It is straightforward that the energy absorption and scattering magnitudes depend strongly on the relative orientation angle $\theta$ between the incident electric field polarization and the AuNP major axis, \textit{i.e.}, $\sigma=\sigma(\theta)$, being $\theta\in[0,\pi/2]$, as can be observed in Fig.\ref{fig:CrossSectionWaveLengthSim} and more clearly in Fig.\ref{fig:CrossSectionWaveTheta}. As expected, the cross-sections values are higher when the electric field of the wave is parallel to the major axis of the AuNP ($\theta=0$), and minimum when it is parallel to the diameter ($\theta=\pi/2$). The laser source used in experiments emits unpolarized light, and the AuNPs are subjected to both random translational and rotational Brownian motions. As a consequence, the actual cross-sections for an individual particle should be compared or estimated by an average, over all possible angles $\theta$, of the calculated cross-sections magnitudes.

\begin{figure}[ht!]
	\centering
	\includegraphics[width=15cm]{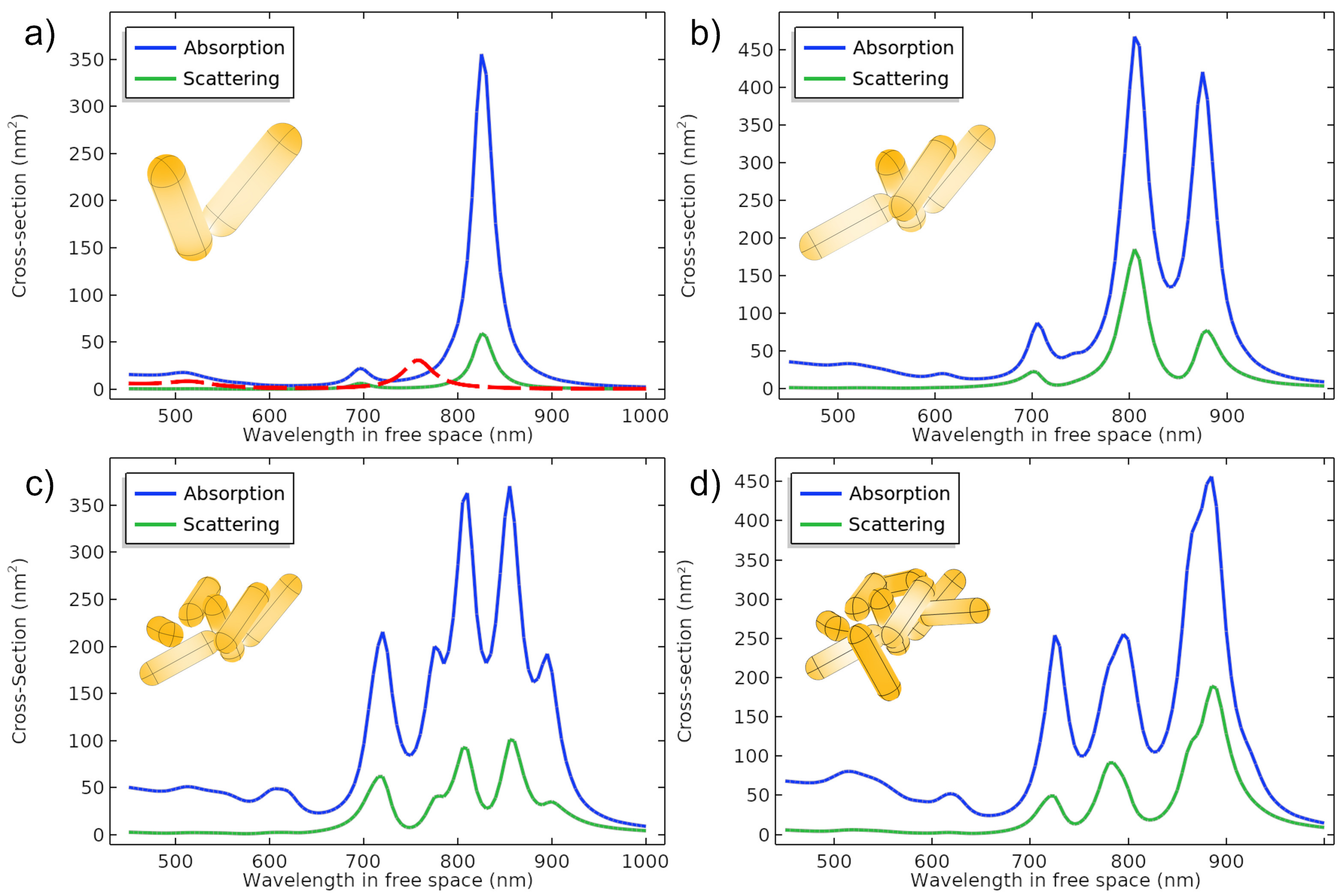}
	\caption{Absorption and scattering cross-sections for random aggregates of a) 2 nanoparticles, b) 4 nanoparticles, c) 6 nanoparticles, and d) 9 nanoparticles. For reference, absorption (red-dashed line) cross-section for an individual nanoparticle has been included in a). Insets: sketches for every aggregate configuration.}
	\label{fig:Aggregates}
\end{figure}

Previous numerical calculations of the laser scattering of different aggregates of nanorods \cite{jain2006calculated,Pratap2022Photothermal} show that the absorption resonance peak position depends strongly on the number of particles of the aggregate and its spatial configuration. The explanation behind this phenomena is that there is a strong interaction between the surface plasmons of neighbouring particles that lead to a red-shift and multiple resonances, even if the aspect ratio of the aggregate with respect to that of a single particle does not change. Similar results were obtained in our own test simulations, see Fig.\ref{fig:Aggregates}, and we have found that the calculated resonance peaks wavelength shift to values much closer to the experimental broad resonance peak (850 nm, Fig.\ref{fig:AbsAu}) than that of an individual nanoparticle (780 nm, Fig.\ref{fig:CrossSectionWaveLengthSim}). Also, we show that the relative magnitude of the resonance peaks depend strongly on the specific configuration of the aggregate.

Another remarkable finding is that, in the case of aggregates, the scattering cross-section is now a larger percentage relative to the absorption cross-section (about 40\% - 60\%), than that found in the case of an individual nanoparticle (a maximum of 14\%, Fig.\ref{fig:CrossSectionWaveLengthSim}). This reflects the occurrence of multiple scattering-absorption events, which would be very unlikely in the case of isolated particles at the considered concentrations in the experiments.

If we consider that our system is a collection of aggregates, with variable number of nanoparticles and random configurations, then the measured extinction spectra should be compared with an average over all aggregation possibilities (in Fig.\ref{fig:Aggregates} there are a few) with weights corresponding to the suspension aggregation statistics. Inspecting the numerical results in Fig.\ref{fig:Aggregates}, it is clear that the weighted superposition of all these possible spectra would result in a curve quite close to the experimental one, Fig.\ref{fig:AbsAu}, with the broad resonance peak at 850 nm.

We then conclude here that: i) there is a certain level of particle aggregation in our system, and that ii) it is not possible to obtain a good theoretical estimate of the absorption cross-section without much more detailed information on the configuration and aggregation statistics. Furthermore, the presence of this aggregation can be observed in the TEM image in the inset of Fig.\ref{fig:AbsAu}, although it is not clear that it may also be due to the TEM preparation itself. Therefore, the effective absorption cross section should be considered as a fitting parameter, as we did in the previous section. The ETP flow simulations have been performed using a fitted value of 2.7·$10^{-17} \text{m}^2$ for the absorption cross-section, being this value in good agreement with the typical range provided by the aggregate simulations for an incident vacuum wavelength of 980 nm.

\section{Conclusions}

We performed a detailed analysis of ETP flow with a novel configuration, namely, with AuNPs dispersed in liquid instead of deposited on a 2D array. We demonstrated that this configuration can lead to strong convection, which could be useful for mixing in microfluidic devices. We evaluated the effects of electric field, laser power and concentration of AuNPs in suspension, finding good agreement with simulations. This comparison allowed us to obtain an estimation of the absorption cross-section of the AuNPs. The obtained value for the absorption cross-section indicates that the AuNPs in suspension are aggregated forming clusters, in agreement with previous experiments. However, we could not get detailed information about their structure, but our results are consistent with clusters formed by 5-9 particles. Such clusters were also observed in TEM images. 

We have demonstrated that the velocity scales quadratically with the concentration of AuNPs. While this dependence should be linear for purely absorbing particles, the presence of aggregates can be invoked to justify this nonlinearity. For this to happen, the scattering and absorption cross-sections should be comparable in magnitude, so events of scattering and absorption become likely in the highest values of concentration. In this situation, absorption is enhanced, leading to an increased ETP flow. This is in agreement with the simulations we performed, since the magnitude of these two cross-sections become comparable in the case of aggregates, while the absorption cross-section is much larger than the scattering one in the case of isolated particles. This nonlinear behavior could be very useful to get very high ETP velocities by inducing some controlled aggregation of the particles.

\section{Methods} 

\subsection{Experimental}

\begin{figure}[ht!]
	\centering
	\includegraphics[width=9.5cm,angle=-90]{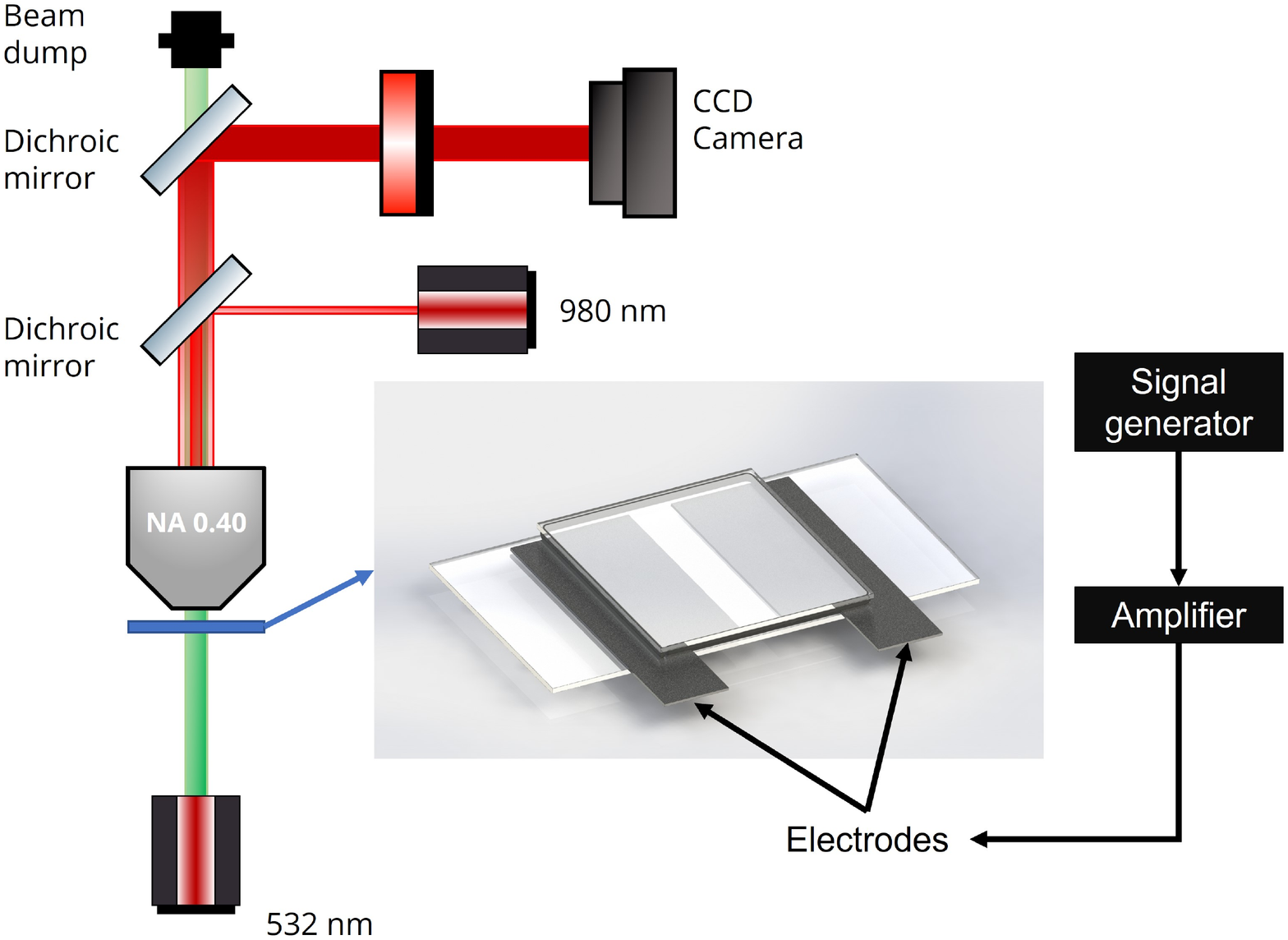}
	\caption{\textbf{Optical setup}. Notice the microfluidic chamber in the centre, as well as the electric scheme on the right. }
	\label{fig:OptSet}
\end{figure}

We built a fluorescence microscope to track the trajectory of microparticles, used as flow tracers. The scheme is shown in Fig.\ref{fig:OptSet}. Briefly, a laser beam ($\lambda=$532 nm) was used to excite the fluorescence of the tracer microparticles (diameter 500 nm, from Sigma Aldrich). A $\lambda=$980 nm laser from Arroyo Instruments was focused to a diffraction-limited spot (diameter 1.56 $\mu$m) incident toward the sample to excite AuNPs, and an electric field was applied with a signal generator from RSPro RSDG 1032X and an amplifier from Falco Systems WMA-100 as can be seen in Fig.\ref{fig:OptSet}

We used one microfluidic cell for each sample, gluing two high-precision coverslips (thickness 170$\pm$5 $\mu$m) with vacuum grease. The mean chamber height was 150$\pm$20 $\mu$m.  During the experiments, we tested different ways to build the microchambers, also using ITO coverslips with 120 $\mu$m spacers. But, this approach changes the direction of the electric field, and also greatly perturbs the heat distribution in the fluid bulk due to the laser heating of the ITO coverslips.

Gold nanoparticles used here are nanorods with 16 nm diameter and are 56 nm long \cite{Arenas2018Electro}. As mentioned before, we worked with five different concentrations. To achieve that, we combined different quantities of gold nanoparticles, Milli-Q water and fluorescent microparticles. Sizing of the AuNPs in suspension by means of dynamic light scattering (Nanosizer Nano ZS, Malvern Instruments) leads to unrealistically small values, likely due to light absorption and localized heating, hence leading to a wrong estimation of their size due to an unknown value of temperature and viscosity around the particles.   

We obtained the experimental flow velocity patterns with PIVLab software \cite{Thielicke2021PIV}, based on the Particle Image Velocimetry technique. For flow visualization, we introduced 500 nm polystyrene fluorescent microspheres from Sigma Aldrich in the samples and recorded the experiments with an IDS UI 1240 camera and a NAVITAR 50 mm lens. In order to achieve a reasonable signal-to-noise ratio, we recorded videos of, at least, 200 frames.

For each experimental condition, one video was recorded; except for the curve shown in fig \ref{fig:EF_LP_Con} c) where two videos were recorded. Each of them were analyzed to obtain a velocity field with an interrogation area of 64 x 64 $\text{pixels}^2$. Averaging out over the frames, we obtain a mean velocity field, in which, we select the maximum value.

\subsection{ETP Simulations}

Additionally, to test the theoretical expression for the ETP driving force \cite{Ndukaife2016Long}, we have performed COMSOL Multiphysics numerical simulations. With these simulations we have solved the strongly coupled heat transfer (Eq.\ref{eq: HeTr}), Navier-Stokes (Eq.\ref{eq: NS1}) and Poisson (Eq.\ref{eq: PO}) equations for calculating temperature, flow velocity and electric fields.
\begin{equation}
\rho c_{\text p}\textbf{u}(\textbf{r})\cdot\nabla T(\textbf{r})-\kappa \nabla^2T(\textbf{r})=q(\textbf{r})
\label{eq: HeTr}
\end{equation}
\begin{equation}
\begin{gathered}
\nabla[p(\textbf{r})\textbf{I}-\mu(\nabla \textbf{u}(\textbf{r})+(\nabla \textbf{u}(\textbf{r}))^T)]=\textbf{F}(\textbf{r}) \\  \nabla\cdot\textbf{u}(\textbf{r})=0
\label{eq: NS1}
\end{gathered}
\end{equation}
\begin{equation}
\nabla^2 \varphi = -\frac{\rho}{\varepsilon}
\label{eq: PO}
\end{equation}
where $\textbf{F}(\textbf{r})$ is the electrothermal force shown in Eq.\ref{eq: ETP}. The main heat source comes from the region of the dispersed nanoparticles illuminated by the heating laser beam. We modelled the heat generated in the AuNPs, using the following expression \cite{Rodriguez2020Heat}:
\begin{equation}
\label{eq: HG}
\textbf{P}_{\text{abs}}=\sigma_{abs} \iint I(\rho,z) \text{Prob}[\rho,z]dS dz
\end{equation}
where we assume that the intensity profile of the laser beam is:
\begin{equation}
\label{eq: HGaux}
I(\rho,z)=I_{inc}\left(\frac{W_{0}}{W(z)}\right)^2 \text{exp}\left(-\frac{2\rho^2}{W^2(z)}\right)
\end{equation}
with $W(z)=W_0\sqrt{1+(z^{}/{z_0})^2}$, $W_0=\sqrt{\lambda z_0^{}/{n\pi}}$ and $z_0=\lambda n^{}/{(\pi(\text{NA}_b)^2)}$ are the beam width, waist radius, and the Rayleigh range, respectively. Also, $I_{inc}$ is the intensity at the center of the focal region ($\rho=z=0$), while NA is the numerical aperture and $n$ is the refractive index of water. Lastly, $\text{Prob}[\rho,z]$ corresponds to the AuNP concentration in the sample, which is assumed to be uniform. In Eq.\ref{eq: HGaux}, we implicitly assume that the relative absorption of light is low, given the fact that the absorption cross-section and the concentration of AuNPs are low. Therefore, the intensity profile is not affected by absorption, and only given by the properties of the focusing system. 

\subsection{Electromagnetic Simulations}

The absorption and scattering cross-section simulations were performed by the use of the finite-elements analysis platform COMSOL Multiphysics, with an Electromagnetic Waves, Frequency Domain physics interface combined with a Wavelength Domain study. The physics interface provides different approaches to compute the electromagnetic fields in the whole system. In this case, we solved only for the perturbation electric field arising from a background plane wave electric field $\textbf{E}_{b}$ incident on the nanoparticles: 
\begin{equation}
\label{eq: BGEF}
\textbf{E}_b = \textbf{E}_0 e^{i \textbf{k} \cdot \textbf{r}}
\end{equation}
where $\textbf{E}_0$ is the incident electric field amplitude, \textbf{k} is the wave number vector in water, and \textbf{r} is the position vector. The physics interface solves the following expression, 
\begin{equation}
\label{eq: EWFD}
 \nabla \times (\nabla \times \textbf{E}) -\kappa^2_0 \varepsilon_r \textbf{E} = 0
\end{equation}
where $\kappa_0$ and $\varepsilon_r$ are the free-space wave number and relative permittivity, respectively.  

The geometry of all electromagnetic simulations, those shown in Figs.\ref{fig:CrossSectionWaveLengthSim}-\ref{fig:Aggregates}, is composed by a sphere representing a certain water domain and, depending on the simulation, one or more gold nanoparticles arbitrarily oriented and distributed close to the center of the sphere. The material of the nanoparticles is gold, with a wavelength-dependent complex refractive index, while the water domain has been simplified to a material model with an only-real part refractive index, with a value of $n_{water}=1.33$. The water domain sphere is surrounded by a spherical shell with a PML (Perfectly Match Layer) setting. This shell has coordinates stretched out toward infinity to mimic an open and non-reflecting infinite domain that absorbs the outgoing waves, and so preventing reflections back into the region of interest. It was also prescribed a Scattering Boundary Condition in the external boundaries of the PML.

From the perturbation electromagnetic fields, we can compute the absorption $\sigma_{abs}$ and scattering $\sigma_{sc}$ cross-sections with the following expressions:
\begin{equation}
\label{eq: sigma_abs}
\sigma_{abs} = \frac{1}{I_0} \int_{V}Q_h=\frac{1}{I_0} \int_{V}\frac{1}{2}\textbf{J}\cdot\textbf{E}^*
\end{equation}
\begin{equation}
\label{eq: sigma_sc}
\sigma_{sc} =\frac{1}{I_0} \int_{S}\textbf{P}_{av}\cdot\textbf{n}
\end{equation}
where $I_{0}$ in the incident intensity, total power dissipation density,\textbf{J} is the electric current density in the nanoparticles, $\textbf{E}^*$ is the complex conjugate of the perturbed electric field, $\textbf{P}_{av}$ is the time averaged perturbation Poynting vector, \textbf{n} is the surface normal, and $V$ and $S$ are the volume and surface of all nanoparticles, respectively.

\section{Acknowledgements}

This research has been supported by FEDER/Junta de Andalucía-Consejería de Economía y Conocimiento/Projects P18-FR-3583 and FQM-410-UGR18, Ministerio de Ciencia, Innovaci\'on y Universidades through Project EQC2018-004693-P, and Grant PID2021-127427NB-I00 funded by MCIN/AEI/ 10.13039/501100011033 and “FEDER Una manera de hacer Europa”. We thank Ángel Delgado and María L. Jiménez for providing the gold nanoparticles and for fruitful discussions.


\providecommand{\latin}[1]{#1}
\makeatletter
\providecommand{\doi}
  {\begingroup\let\do\@makeother\dospecials
  \catcode`\{=1 \catcode`\}=2 \doi@aux}
\providecommand{\doi@aux}[1]{\endgroup\texttt{#1}}
\makeatother
\providecommand*\mcitethebibliography{\thebibliography}
\csname @ifundefined\endcsname{endmcitethebibliography}
  {\let\endmcitethebibliography\endthebibliography}{}

\end{document}